\documentclass[aps,prb,final,twocolumn,showpacs]{revtex4}

\usepackage{graphicx}
\usepackage{dcolumn}
\usepackage{bm}

\usepackage{color}

\begin{document}

\title{Vortex and gap generation in gauge models of graphene}

\author{O. Oliveira$^{*,\ddagger}$, C. E. Cordeiro$^\dagger$, A. Delfino$^\dagger$,  W. de Paula$^*$ and T. Frederico$^*$}

\affiliation{$^\dagger$Instituto de F\'isica, Universidade Federal Fluminense, 24210-3400- Niter\'oi - RJ, Brazil \\
                 $^*$Departamento de F\'{\i}sica, Instituto Tecnol\'ogico de Aeron\'autica,
                               12.228-900, S\~ao Jos\'e dos Campos, SP, Brazil \\
                 $^\ddagger$Departamento de F\'{\i}sica, Universidade de Coimbra, 3004-516 Coimbra, Portugal}

\date{\today}

\begin{abstract}
Effective quantum field theoretical continuum models for graphene are investigated. The models include a
complex scalar field and a vector gauge field. Different gauge theories are considered and their gap patterns
for the scalar, vector, and fermion excitations are investigated. Different gauge groups lead to different relations
between the gaps, which can be used to experimentally distinguish the gauge theories. In this class of models the
fermionic gap is a dynamic quantity. The finite-energy vortex solutions of the gauge models have the flux of the
"magnetic field" quantized, making the Bohm-Aharonov effect active even when external electromagnetic fields
are absent. The flux comes proportional to the scalar field angular momentum quantum number. The zero modes
of the Dirac equation show that the gauge models considered here are compatible with fractionalization.
\end{abstract}


 \maketitle

\section{Introduction and Motivation \label{introducao}}

In a monolayer of graphene \cite{wallace1947,novoselov2004,abergel2007,blake2007,casiraghi2007},
the single-particle dispersion relation near the so-called $K$ and $K^\prime$ Dirac points is linear in
$|\vec{k}|$ -- see, for example, \cite{castro2009,Peres2010}.
Formally, it is the dispersion relation of a massless relativistic fermion.
Furthermore, the description of the low energy electronic excitations can be
accommodated in a Dirac-type equation. Indeed, starting from a tight-binding
Hamiltonian, with a hopping parameter independent of the lattice site,
one can compute exactly the dispersion relation, expand around the two
inequivalent Dirac points to rewrite the dynamical equations of motion for
electrons and holes as a Dirac equation in two dimensions with the four
component spinor given by
\begin{equation}
 \Psi ~ =  ~ \left( \begin{array}{c}
\psi^b_+ \\ \psi^a_+ \\ \psi^a_- \\ \psi^b_- \end{array} \right).
\end{equation}
The indices $a$ and $b$ refer to the two triangular sublattices,
while the $+$ and $-$ indices to the two Dirac $K$ and $K^\prime$
points. For a perfect graphene crystal structure the fermions behave
as massless relativistic particles \cite{novoselov2005}, which
translates into the well known ballistic behavior of the electrons
\cite{novoselov2004b,Morozov2005,Du2008}, and there is no gap
between valence and conduction bands. However, if the two
dimensional honeycomb array of carbon atoms is distorted due to the
presence of impurities or to the distortion of the crystal
structure, for example, the fermions acquire an effective gap given
by half of the mass gap. Besides fermion mass generation, quantum
Hall effect
\cite{novoselov2005,zhang2005,Gusynin2005,peres2006,Peres2006a,zhang2006,zhao2010},
fractionalization\cite{Hou_et_al_2007,Jackiw_Pi_2007,chamon2008} and
Berry phases\cite{zhang2005} have been observed in two dimensional
graphenelike structures.

Fractionalization in one dimensional models was investigated more than three decades ago
\cite{Jackim_Rebbi_1976,Su1979,jackiw1981} within polyacetylene. A similar phenomena like fractional
quantum hall effect due to quasiparticle fractional charge and/or fractional statistics can take place in two
dimensional systems\cite{zhang2005,zhang2006,zhao2010}.

A dynamical theory for two dimensional graphene should describe, of course, its
phenomenology and should be able to accommodate for the possibility of gap generation and
fractionalization. In \cite{Hou_et_al_2007} the authors presented a mechanism for
electron fractionalization in graphenelike systems keeping time reversal symmetry.
Invoking a Kekul\'e texture, a complex order parameter $\Delta_0$ was introduced.
$\Delta_0$ couples the two Dirac points and changes the electron dispersion relation to
$\epsilon(\vec{k}) = \sqrt{ \vec{k}^2 + |\Delta_0|^2}$. Assuming a vortex-like profile
$\Delta(\vec{r}) = \Delta(r) \, e^{i n \theta}$, where $n$ is
an integer, with $\Delta(r)$ vanishing as $r^{|n|}$ for small r and approaching $\Delta_0$ at large $r$,
fractionalization was associated with the presence
of a zero mode of the Dirac kernel. Although fractionalization was connected with the vortex shape of
$\Delta(\vec{r})$, in
\cite{Hou_et_al_2007} the authors do not specify the dynamics for the complex vortex profile.

In \cite{Jackiw_Pi_2007} a dynamical content to the vortices was introduced through
a chiral gauge theory which is compatible with electron fractionalization, extending
the work of Hou \textit{et al} \cite{Hou_et_al_2007}. In the language of \cite{Jackiw_Pi_2007},
vortices are associated with a complex scalar field $\varphi$ which couples linearly to the fermions.
Although, a dynamical equation was written, see equations (13) and (14) in \cite{Jackiw_Pi_2007},
the potential $V(\varphi^* \varphi)$ was left unspecified.
From the point of view of Quantum Field Theory (QFT), there is no reason to exclude other types of
symmetries and couplings, not present in the model discussed by Jackiw and Pi, without
destroying fractionalization.

Indeed, field theoretical models have been applied to describe
nanotubes and graphene physics with some success in reproducing
their quantum properties (see e.g. \cite{Cordeiro,chaves}). In this
work, we elaborate on derivative free fermion-boson and boson self
interactions allowed by QFT principles and discuss possible gauge
interactions. The models considered here are a generalization of the
results of \cite{Jackiw_Pi_2007} and, besides the fermionic field,
they consider a complex scalar field $\varphi$ and a single gauge
vector $A_\mu$ field. Moreover, possible ways to distinguish between
the different gauge symmetries are discussed.

Our interpretation for the complex scalar field and gauge field being that
$\varphi$ and $A_\mu$ resume all the dynamics of the self-interaction of the carbon background
and the mean fermionic self interaction.

The usage of scalar and vector potentials to describe some of the graphene properties
is not new. Indeed, scalar and vector potentials, including gauge fields, have been used in the
literature to describe disorder phenomena, including distortions of the lattice honeycomb,
structural defects, point defects and self-doping effects associated with the breaking of
electron-hole symmetry near the Dirac points among other properties.
A detailed discussion can be found in \cite{castro2009,Voz2010} and references there in.

Graphene is an electrical neutral system. On the other hand graphene has charge carriers.
Therefore, it seems natural to associate a charged field with the carbon background.
Furthermore, if $\varphi$ resumes the carbon background self-interactions it should be able to
accommodate for the propagation of phonons in the carbon lattice. Phonons feel the density of states
of the fermionic degrees of freedom and one expects $\varphi$ to couple to
the density of electron/holes, i.e. to
$\overline \psi \, \psi =
- \left( \psi^b_+ \right)^\dagger \psi^a_-  -
\left( \psi^a_+ \right)^\dagger \psi^b_-   -
\left( \psi^a_- \right)^\dagger \psi^b_+  -
\left( \psi^b_- \right)^\dagger \psi^a_+ \, .
$
Throughout this paper, we will use the chiral representation for the Dirac matrices, where
\begin{equation}
 \gamma^0 = \left( \begin{array}{ll} 0 & -1 \\  -1 & 0 \end{array} \right), \qquad
 \vec{\gamma} = \left( \begin{array}{ll} 0 & \vec{\sigma}  \\  - \vec{\sigma} & 0 \end{array} \right),
\end{equation}
and
\begin{equation}
 \gamma_5 = \left( \begin{array}{ll} 1 & 0  \\  0 & -1 \end{array} \right);
\end{equation}
$\sigma^j$ stand for the Pauli matrices.
Besides the coupling to the density of electron/holes, QFT allows also for a pseudoscalar like interaction,
described by a coupling of $\varphi$ to
$\overline \psi \,\gamma_5 \psi =
 \left( \psi^b_+ \right)^\dagger \psi^a_-  +
\left( \psi^a_+ \right)^\dagger \psi^b_-   -
\left( \psi^a_- \right)^\dagger \psi^b_+  -
\left( \psi^b_- \right)^\dagger \psi^a_+ \, .
$
The scalar and pseudoscalar interactions couple
the two triangular sublattices and the two Dirac points $K$ and $K^\prime$
in different ways.
The models discussed in the present work explore
contributions coming from both type of interactions (i.e.,
 $\overline\psi \, \psi$ and $\overline\psi \gamma_5 \psi$).

The potential energy for the complex scalar field $\varphi$ can
accommodate a non-vanishing vacuum expectation value $\langle
\varphi \rangle$. If $\langle \varphi \rangle \ne 0$, then the model
generates a fermion mass via spontaneous symmetry breaking. On the
other hand, if $\langle \varphi \rangle = 0$, the electrons in
graphene remain gapless. Therefore, we identify pure graphene with
the vacuum state where $\langle \varphi \rangle = 0$, with all other
graphene distorted and/or doped states begin described by different
vacuum and for these $\langle \varphi \rangle \ne 0$.

In what concerns the bosonic fields, the model can accommodate mass gaps both for the
scalar and vector excitations. In general, for the gauge theories considered here, a fermion mass
gap implies also a vector mass gap. This comes directly from the Higgs mechanism for mass generation.
The gap for the scalar excitations is linked with the details of the potential $V(\varphi^\dagger\varphi)$
and is not directly coupled with the fermion and vector gaps.
Indeed, we found that the scalar gap can vanish independently of
the fermion and vector gaps.

In this paper we also discuss a number of different gauge models
which, in principle could be suitable to describe  graphene
properties. The relation between the spectrum of the scalar and
vector excitations with the fermionic spectra depends on which
symmetry is gauged. Further, the different connections between
fermion, scalar and vector gaps opens for the possibility to check
experimentally which of the gauge symmetries, if any, is realized in
graphene. Changing the fermion mass gap, for example modifying the
concentration of impurities and or the distortion of the lattice,
and looking on how the scalar/vector mass excitations adjust
themselves, one can distinguish between the different gauge models.
 Besides the pattern of the mass gaps, in general,
the models allow also for vortex like solutions and are compatible
with fractional statistics.

These gauge models have finite energy vortex solutions. For one
example, we show that the vortex solution implies in the
quantization of the ``magnetic field" flux. In this case, the
abelian gauge field is connected with the angular momenta of
$\varphi$ along an axis perpendicular to the graphene sheet and, in
this sense, the flux of the ``magnetic field" is a measure of the
angular momenta of $\varphi$. This particular solution can be
interpreted as consequence of topological defects in the graphene
structure and, in principle, phenomena like the Bohm-Aharonov
\cite{ehrenberg1949,aharonov1959,aharonov1961} effect can occur even when external
electric and magnetic fields are absent.

We show that the vortex solutions of non-chiral gauge models
presented here, have normalizable zero modes of the Dirac equation.
The presence of the normalizable zero modes implies
fractionalization for graphene and the quantum Hall effect in
graphene sheets could become possible even without chiral gauge
symmetry and without external electric and magnetic fields.
The observation of the quantum Hall effect in two dimensional materials
without external electromagnetic fields was also discussed, within the framework
of tight binding models, in the work of Haldane \cite{haldane1988} and
Hill \textit{et al}\cite{hill2010}.
According to the later work, the observation of the quantum Hall effect
without an electromagnetic field requires
the breaking of the sublattice symmetry, where the two sublattices $a$ and
$b$ are interchanged, and the opening of a mass gap at one of the Dirac points,
let us say $K$, while the other Dirac point $K^\prime$ remains gapless. In the class
of models discussed here, the mass gap is open, or not, simultaneously at $K$ and
$K^\prime$. Then, according to \cite{hill2010} quantum Hall effect without an
external magnetic field is not measurable as the contributions from $K$ and $K^\prime$
to the Hall conductivity cancel exactly. The models considered in the present work,
although reproduce the tight binding model in the appropriate limit
(see, for example, the paper\cite{castro2009}) they give dynamics to all
the fields that represent the carbon graphene background and the excitations
$\varphi$ and the gauge field.
Recall that $\varphi$ is charged and can give rise to an electric current.
In this sense, the models go beyond the tight binding model, opening the possibility
of having a dynamical situation where the conditions explored in \cite{hill2010} do no apply,
and, perhaps, may allow the measurement of the quantum Hall effect in graphene. We plan
to address this question in a future publication.

We would like to call the reader attention that, within the class of gauge models
discussed in the present work, fractionalization is allowed without
breaking any of the usual discrete symmetries like, for example,
time reversal. We do not compute the rich set of zero modes of the
Dirac equation but those ones obtained here are, again, connected
with the angular momentum of the complex scalar field $\varphi$.

This paper is organized as follows. In section \ref{the_model} the
effective QFT for graphene are discussed, including its global
symmetries. In section \ref{the_gauges}, the different global
symmetries are gauged and we discuss how this change the spectra of
the scalar, vector and fermion excitations. Furthermore, combining
the information on the different types of gaps, we are able to
suggest an experimental test to disentangle which of the gauge
symmetries apply to graphene. In section \ref{eq_movimento} the
equations of motion are derived and the vortex solutions for the
gauge models are discussed. The  short distance and long distance
properties of the vortex are computed explicitly. The gauge models
predict the flux quantization of the ``magnetic field" associated
with the gauge field. Further, the flux quantization is connected
with the angular momenta of $\varphi$. In section
\ref{fractionalizacao} the zero mode solutions of the Dirac equation
for a vortex configuration are investigated. Finally, in section
\ref{fim} we resume and conclude.

\section{The Effective Model \label{the_model}}

Let us assume that the charge carries, i.e. electrons and holes, are relativistic fermions described by
a four component spinor $\psi$.
The Lagrangian density describing the interaction between fermions $\psi$ and $\varphi$ can be written as
\begin{eqnarray}
  \mathcal{L}  & =  & \overline\psi  i \, \gamma^\mu \partial_\mu \psi
  +  \partial^\mu \varphi^\dagger \partial_\mu \varphi \nonumber \\
  & &  \qquad
                        \, - \,   P(\varphi) \, \overline\psi \psi
                        \, - \, P_5(\varphi ) \, \overline \psi \gamma_5 \psi
  - \, V( \varphi^\dagger \varphi ) \, ,
  \label{new_lagrangian}
\end{eqnarray}
where the polynomial $P( \varphi)$ and $P_5( \varphi)$ define the type of interaction between fermions and the
carbon crystal structure and $V(\varphi^\dagger \varphi)$ the self interactions of the background structure.

The reader should note the linear combination of $P \, \overline\psi
\psi$ and $P_5 \, \overline\psi \gamma_5\psi$ couplings. Such a
freedom would allows to set different couplings to each of the
possible fermion chiralities and, in this way, build a chiral
theory. Moreover, other Dirac $\gamma$-matrices are allowed.
However, to keep it as simple as possible and to avoid derivative
couplings, in the following we will consider only scalar and
pseudoscalar types of interactions.

In a system of units where the action is dimensionless, space and time have dimensions of an inverse mass,
and for two spatial dimensions and one temporal dimension, $\psi$ has dimension of mass,
$[ \psi ] \sim M$,  and $\varphi$ of square root of mass,
$[ \varphi ]�\sim M^{1/2}$.
Requiring that the theory described by $\mathcal{L}$ is perturbatively renormalizable, then for polynomial
interactions, naive power counting
forbids coupling constants $[ g ] \sim M^\alpha$ with $\alpha < 0$. This fixes unambiguously the interaction
terms to
\begin{equation}
   P(\varphi) = g_1 \left( \varphi + \varphi^\dagger \right) ~ + ~ g_2 \, \varphi^\dagger \varphi \, ,
   \label{fermiao_phi}
\end{equation}
\begin{equation}
   P_5( \varphi) =  h_1 \left( \varphi - \varphi^\dagger \right) ~ + ~ i \, h_2 \, \varphi^\dagger \varphi \, ,
   \label{fermiao_pseudo_phi}
\end{equation}
and
\begin{equation}
   V(\varphi^\dagger \varphi) ~ = ~ \mu^2 \left( \varphi^\dagger\varphi \right) ~ + ~ \frac{\lambda_4}{2} \left( \varphi^\dagger\varphi \right)^2
                          + ~ \frac{\lambda_6}{3} \left( \varphi^\dagger\varphi \right)^3
  \label{potential_phi}
\end{equation}
up to a constant $V_0$.

If in $P(\varphi)$ and $P_5 ( \varphi )$ one takes $h_1 = -g_1$ and $g_2=h_2=0$ one recovers
the Jackiw-Pi theory with their
$\varphi^r = 2 \, \mbox{Re}(\varphi)$ and $\varphi^i = 2 \, \mbox{Im}(\varphi)$ -- see equation (8) in
 \cite{Jackiw_Pi_2007}. In this sense $\mathcal{L}$ generalizes the results of \cite{Jackiw_Pi_2007}.

Let us discuss now the global symmetries of the model described by
the Lagrangian density (\ref{new_lagrangian}).

\subsection{Global $U_A(1)$ Symmetry}

One of the motivations of \cite{Jackiw_Pi_2007} was to build a chiral gauge theory.
So let us consider the same type of chiral transformation, i.e.
\begin{equation}
  \psi \longrightarrow e^{i \omega \gamma_5} \,\psi \, , \hspace{0.7cm}
   \varphi \longrightarrow e^{i \eta} \, \varphi \, .
   \label{chiral_transformation}
\end{equation}
To first order in $\omega$ and $\eta$, the corresponding variation of the Lagrangian density reads
\begin{widetext}
\begin{equation}
 \Big\{ 2 i \omega
           \left[ h_1 \left( \varphi - \varphi^\dagger \right) + i h_2 \left( \varphi^\dagger \varphi \right) \right] +
            i \eta g_1 \left( \varphi - \varphi^\dagger \right)  \Big\} \, \overline\psi \psi ~ + ~
 \Big\{ 2 i \omega
           \left[ g_1 \left( \varphi + \varphi^\dagger \right) + g_2 \left( \varphi^\dagger \varphi \right) \right] +
            i \eta h_1 \left( \varphi + \varphi^\dagger \right)  \Big\} \, \overline\psi \gamma_5 \psi \, .
\end{equation}
\end{widetext}
Requiring invariance of $\mathcal{L}$ under the transformation (\ref{chiral_transformation}), it follows that
$g_2 = h_2 = 0$, as in the Jackiw-Pi theory, and $g_1 = \pm h_1$ and $\eta = \pm 2 \omega$, with
the minus sign recovering the original Jackiw-Pi theory. Note that,
from the point of view of the  (\ref{chiral_transformation}), invariance of the theory
means that the chiral charge associated with $\varphi$ is, up to a sign, twice the fermionic charge.

The set of transformations (\ref{chiral_transformation}) with  $\eta = \pm 2 \omega$
form a group which will be called from now on $U_A(1)$. Recall that the Lagrangian density is invariant under
$U_A(1)$ if and only if $g_2 = h_2 = 0$ and $g_1 = \pm h_1$.

\subsection{Global $U(1)$ Symmetries}

Besides the chiral transformation just discussed, the Lagrangian density (\ref{new_lagrangian}) has
further non-chiral $U(1)$ global symmetries. The set of transformations
\begin{equation}
  \psi \longrightarrow e^{i \omega} \,\psi \,
   \label{Uf_transformation}
\end{equation}
defines the $U_f(1)$ global symmetry of $\mathcal{L}$ and the set
\begin{equation}
  \varphi \longrightarrow e^{i \omega} \,\varphi \, ,
   \label{Ub_transformation}
\end{equation}
defines the $U_b(1)$ global symmetry of  $\mathcal{L}$ if $g_1 = h_1 = 0$. Further, if $g_1 = h_1 = 0$
in (\ref{fermiao_phi}) and (\ref{fermiao_pseudo_phi}), then the set of transformations
\begin{equation}
  \psi \longrightarrow e^{i \omega} \,\psi \, , \hspace{0.7cm}
   \varphi \longrightarrow e^{i \eta} \, \varphi \, ,
   \label{Ufb_transformation}
\end{equation}
where $\omega$ and $\eta$ are independent parameters
defines another global symmetry of $\mathcal{L}$, called below $U_f(1) \otimes U_b(1)$. Gauging this
symmetry requires the introduction of two gauge fields unless one imposes an additional discrete
symmetry with respect to the interchange of the gauge fields. Note that the discrete symmetry gives
no constraint on the coupling constants for the fermionic, $g$, and bosonic, $g_\varphi$, fields.
Indeed, the fermionic covariant derivative reads $D_\mu = \partial_\mu + i g A_\mu$,
while the bosonic covariant derivative is $D_\mu = \partial_\mu + i g_\varphi A_\mu$, where
$A_\mu$ is the gauge field. For the sake of simplicity, i.e. to avoid
considering more than one gauge field, we will analyze only the $U_f(1) \otimes U_b(1)$
symmetry supplement with the discrete symmetry. Anyway, we will keep using
the name $U_f(1) \otimes U_b(1)$ for the symmetry.

\subsection{On The Various Global Symmetries}

The various global symmetries are distinguished by the nature of the $\varphi - \psi$ interaction and
by the number of independent coupling constants associated with the gauge field required to
define the model.

The $U_A(1)$ symmetry is not compatible with the interactions
$\left( \varphi^\dagger \varphi \right) \, \overline\psi \, \psi$ and
$\left( \varphi^\dagger \varphi \right)\, \overline\psi \,\gamma_5 \psi$ and
only linear in $\varphi$ terms are allowed in the interaction with fermions. Further, the model defines a unique
coupling constant.

The $U_f(1)$ allows linear and quadratic $\varphi$ couplings to the fermionic field and requires a unique
gauge coupling constant.

The local symmetries $U_b(1)$ and $U_f(1) \otimes U_b(1)$ are compatible only with a
quadratic $\varphi$ coupling to the fermion fields. If  $U_b(1)$ requires a unique gauge coupling constant,
the gauge model with $U_f(1) \otimes U_b(1)$ as a symmetry group includes two independent gauge
couplings.

Table \ref{tab_sumario_simetrias} resumes the global symmetries of Lagrangian density
(\ref{new_lagrangian}) and the corresponding constraints on the $\varphi - \psi$ coupling constants.

\begin{table}[t]
   \centering
   \begin{tabular}{c @{\hspace{0.6cm}} l } 
      \toprule
      Symmetry    & Constraints \\
      \hline
      $U_A(1)$    & $g_2 = h_2 = 0$, $g_1 = \pm h_1$ \\
      \hline
      $U_f(1)$      & none  \\
      $U_b(1)$     & $g_1 = h_1 = 0$  \\
      $U_f(1) \otimes U_b(1)$      & $g_1 = h_1 = 0$  \\
      \toprule
   \end{tabular}
   \caption{Global symmetries of the Lagrangian density (\ref{new_lagrangian}). Recall that
                  for $U_f(1) \otimes U_b(1)$ we impose an additional discrete symmetry -- see text
                  for discussions.}
   \label{tab_sumario_simetrias}
\end{table}

\section{Gauge Models and Mass Gap \label{the_gauges}}

The various $U(1)$ symmetries of $\mathcal{L}$, see  table  \ref{tab_sumario_simetrias},
can be made local. Different symmetries will lead to different gauge theories for graphene, after
replacing the derivatives by covariant derivatives and adding the corresponding kinetic term
for the gauge field. Naturally, the different symmetries will introduce different dynamics which can
be seen, for example, at the level of the theory spectra, i.e. at the various mass gaps.

Recall that we are excluding derivative type couplings. In what concerns the gauge field, not including
derivative couplings means that \textit{a priori} we are excluding a Chern-Simons term
\cite{chern1974,deser1982,witten1989}
\begin{equation}
   \epsilon^{\alpha\beta\gamma} A_\alpha \left( \partial_\beta A_\gamma\right)
\end{equation}
in $\mathcal{L}$. This type of interaction is allowed by gauge invariance and, in 2+1 dimensions, is not excluded by
the renormalizability requirement.

\subsection{Scalar Mass Gap}

We start our discussion looking at the scalar excitations in graphene, i.e. looking at
the mass spectra for the complex scalar field $\varphi$.

The self interactions of $\varphi$ are described by the potential energy $V(\varphi^\dagger\varphi)$ --
see equation (\ref{potential_phi}).
Depending on the values for $\mu^2$, $\lambda_3$ and $\lambda_6$, $V$ can have either one,
two or three minima. The discussion of the $V(\varphi^\dagger \varphi)$ extrema is
relatively straightforward and will not be reproduced here. The relation between potential parameters and
number of extrema is summarized in table \ref{tab_sumario_extremos}.

The mass gap for the scalar excitations, i.e. the mass associated with the complex scalar field, can be
computed from (\ref{potential_phi}) writing $\varphi = v + \Phi$, where $v = \langle \varphi \rangle$ is the
vacuum expectation value of $\varphi$ assumed to be real. If $\varphi$ can be rotated in such a way that
it becomes a real field, then it follows that the quadratic term in $V(\Phi^2)$ is given by
\begin{equation}
\frac{1}{2} \, M^2_\Phi \, \Phi^2 = 2 v^2 \left( \lambda_4 + 2 v^2 \lambda_6\right) \, \Phi^2
\label{mphi}
\end{equation}
and one can define the mass gap for scalar excitations as
\begin{equation}
\Delta_\Phi =  M_\Phi = 2 \, |v| \, \sqrt{ \lambda_4 + 2 v^2 \lambda_6} \, .
\label{deltaphi}
\end{equation}
The scalar mass gap is, then, independent of the gauged symmetry.

A non-vanishing $M_\Phi$
requires either a non-vanishing expectation value for $\langle \varphi \rangle$ or a
$\langle \varphi \rangle = 0$ and a $\mu^2 > 0$. Furthermore, a non-vanishing scalar gap requires also
that $\lambda_4 + 2 v^2 \lambda_6 > 0$, if $\langle \varphi \rangle \ne 0$.

From equations (\ref{mphi}) and (\ref{deltaphi}) it follows that the model is compatible with
a non-vanishing $\varphi$ vacuum expectation value, i.e. a $\langle \varphi \rangle \neq 0$,
in combination with a vanishing scalar gap if and only if $\lambda_4 = - 2 v^2 \lambda_6$.
In this case, the theory predicts a fermionic mass gap proportional to $\langle \varphi \rangle$,
with no gap formation for the scalar excitations, i.e the dispersion relation for the scalar excitations is
linear in $| \vec{k}|$.

\begin{table}[t]
   \centering
   \begin{tabular}{c @{\hspace{0.6cm}} c @{\hspace{0.6cm}} c @{\hspace{0.6cm}} c @{\hspace{0.9cm}} r} 
      \toprule
      \multicolumn{2}{c}{Item} \\
      $\mu^2$    & $\lambda_4$ & $\lambda_6$ & $\delta$ & \# extrema\\
      \hline
      $< 0$         & $> 0$  & $< 0$ & $0$ & 1 maximum  \\
      $> 0$         & $<0$  & $> 0$ & $0$ & 1 minimum  \\
      \hline
      $> 0$         & $> 0$  & $> 0$ & $> 0$ & 1 minimum  \\
      $> 0$         & $< 0$  & $> 0$ & $> 0$ & 5 extrema \\
      $< 0$         & any     & $> 0$ & $> 0$ & 3 extrema \\
      \toprule
   \end{tabular}
   \caption{The number of extrema of $V(\Phi)$ as a function of the potential parameters. Our definition
                for $\delta$ is $\delta = \lambda^2_4 - 4 \, \mu^2 \lambda_6$.}
   \label{tab_sumario_extremos}
\end{table}

\subsection{Fermionic Mass Gap}

For the fermions fields, if  $\varphi$ and/or $\varphi^\dagger \varphi$ acquire a non-vanishing vacuum
expectation value, then $\mathcal{L}$ acquires a mass term and a chiral mass term -- see
equations (\ref{new_lagrangian}), (\ref{fermiao_phi}) and (\ref{fermiao_pseudo_phi}).

Let us assume that $\langle \varphi \rangle = v \ne 0$, with $v$ being a real number.
From the point of  view of the fermions themselves, the interaction with the carbon structure shows up as
\begin{equation}
   \widetilde{m} \, \overline\psi \, \psi  ~ + ~ i \, h_2 \, v^2 \,\overline\psi \, \gamma_5  \, \psi \,
   \label{mass_term}
\end{equation}
where $\widetilde{m} = 2 \, g_1 \, v + g_2 v^2$.
There is no reason \textit{a priori} to require the positivity of $\widetilde{m}$ or $h_2$.
Indeed, solving the free Dirac equation, with a mass term given by  (\ref{mass_term}), gives the following
dispersion relation
\begin{equation}
 \epsilon (\vec{p}^{~ 2}) = \sqrt{\vec{p}^{~ 2} + \widetilde{m}^2 + h^2_2 v^4} \, ,
\end{equation}
i.e. the effective fermion mass is given by $\sqrt{\widetilde{m}^2 + h^2_2 v^4}$ and
the corresponding mass gap
between valence and conducting bands is $2 \, \sqrt{\widetilde{m}^2 + h^2_2 v^4} ~$.
The above reasoning is valid even when $\langle \varphi \rangle  = 0$ and
$\langle \varphi^\dagger \varphi \rangle \ne 0$.
In this case $\widetilde{m} = g_2 \langle \varphi^\dagger \varphi \rangle$ and
$h_2 \langle \varphi^\dagger \varphi \rangle$ replaces $h_2v^2$.

The coupling of the complex scalar field $\varphi$ to the fermion degrees of freedom is given by
\begin{eqnarray}
 & &
  \Big( g_1 \left( \varphi + \varphi^\dagger \right) ~ + ~ g_2 \, \varphi^\dagger \varphi \Big) ~ \overline\psi \, \psi
   \nonumber \\
  & &
  \quad + ~\Big( h_1 \left( \varphi - \varphi^\dagger \right) ~ + ~ i \, h_2 \, \varphi^\dagger \varphi \Big) ~
  \overline\psi \, \gamma_5 \psi \, ,
\end{eqnarray}
which has exactly the same structure as the mass term given by equation (\ref{mass_term}). Therefore,
the status of the field $\varphi$ can be translated into a dynamical fermion mass, i.e. a dynamical mass gap,
which is both time and spatial dependent. The model accommodates graphene states
where for certain space-time regions the system is gapless, i.e $\varphi \ne 0$, and for others where
$\varphi = 0$ and there is no gap. We are currently
exploring the implications of this dynamical gap to graphene properties and will report the results elsewhere.

In graphene the fermionic mass gap $\Delta$ is a function of gauged symmetry.
It follows that for
\begin{eqnarray}
 & U_A (1)  & \hspace{0.5cm}  \Delta = 4 \left| g_1 v \right| \, , \\
 &  U_f (1)  & \hspace{0.5cm}  \Delta = 2 \sqrt{\widetilde{m}^2 + h^2_2 v^4} \, , \\
 &   U_b (1), ~  U_f (1)   \otimes U_b (1)  & \hspace{0.5cm}  \Delta = 2 \sqrt{ g^2_2 + h^2_2} \, v^2
\end{eqnarray}
where $\Delta$ is twice the fermion mass.

\subsection{Vector Mass Gap}

It remains to discuss the mass gap for the vector excitations in graphene.
The mass term for $A_\mu$ is generated by the scalar kinetic part of $\mathcal{L}$, i.e. by
$\left( D_\mu \varphi \right)^\dagger D^\mu \varphi$. Therefore, unless the gauge transformation
changes $\varphi$, the gauge field remains massless. It follows that for
\begin{eqnarray}
 & U_A (1)  & \hspace{0.5cm}  \Delta_A = \sqrt{2} \, \left| g_1 \, v \right| \, , \\
 &  U_f (1)  & \hspace{0.5cm}  \Delta_A = 0 \, , \\
 &   U_b (1), ~  U_f (1)   \otimes U_b (1)  & \hspace{0.5cm}  \Delta_A = \sqrt{2} \, \left| g_\varphi \, v \right|
\end{eqnarray}
where $\Delta_A$ is the vector mass, i.e. the vector mass gap.

\subsection{Gauge Symmetries and Gap Relations \label{seccao_mass_gap}}

\begin{table}[t]
   \centering
   \begin{tabular}{c @{\hspace{0.6cm}} l @{\hspace{0.6cm}} l }
      \toprule
      Symmetry    & $\Delta$ & $\Delta_A$ \\
      \hline
      $U_A(1)$    & $4 \, \left|g_1 \, v \right|$  & $\sqrt{2} \, \left| g_1 \, v \right|$ \\
      \hline
      $U_f(1)$      & $2 \sqrt{\widetilde{m} + h^2_2 v^4}$ & 0 \\
      $U_b(1)$    &  $2 \sqrt{g^2_2 + h^2_2 } ~ v^2$ & $\sqrt{2} \, \left| g_\varphi \, v \right|$ \\
      $U_f(1) \otimes U_b(1)$   &  $2 \sqrt{g^2_2 + h^2_2 }\, v^2$ &  $\sqrt{2} \, \left| g_\varphi \, v \right|$ \\
      \toprule
   \end{tabular}
   \caption{Mass gaps as a function of the gauge symmetry -- see text
                  for discussions. The scalar mass gap is independent of gauge group and
                  for a non-vanishing vacuum expectation value if given by
                  $\Delta_\Phi = 2 \, |v| \, \sqrt{ \lambda_4 + 2 v^2 \lambda_6 }$.}
   \label{tab_mass_gaps}
\end{table}

The mass, i.e. the gaps, for each of the fields in the model are generated via Higgs mechanism.
Besides the mass, the Higgs mechanism also provides a relation, dependent on the symmetry group,
between the different gaps - see Table \ref{tab_mass_gaps}  for a summary of the results
discussed in the previous sections.

A $U_A (1)$ chiral gauge theory implies a linear relation between
the vector and fermion mass gaps $\Delta_A = \Delta/\sqrt{8}$, while
$U_b(1)$ or $U_f(1) \otimes U_b(1)$ relate the two mass gaps by a
quadratic equation $\Delta^2_A = \Delta ~ g^2_\varphi / \sqrt{g^2_2
+ h^2_2}$. For the gauge theory associated with the $U_f (1)$
symmetry, there is a mass gap for the fermionic and scalar degrees
of freedom, but no gap for the vector excitations.

A non-vanishing fermionic mass gap requires a $\langle \varphi \rangle \ne 0$ or
$\langle \varphi^\dagger \varphi \rangle \ne 0$, which by itself implies
a scalar mass gap in graphene. The connection between the scalar mass gap $\Delta_\varphi$ and
the remaining gaps is slightly more complicated than the relation between $\Delta$ and $\Delta_A$.

\begin{figure}[t] 
   \vspace{0.6cm}
   \centering
   \includegraphics[width=3in]{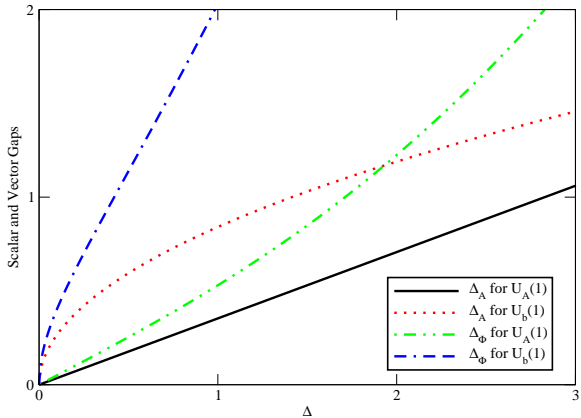}
   \caption{$\Delta_A$ and $\Delta_\Phi$ as function of the fermionic mass gap $\Delta$. The curves where
                 computed setting all the coupling constants and the potential parameters to one and using arbitrary units.}
   \label{fig_mass_gap_delta}
\end{figure}

The scalar, fermion and vector mass gaps are functions of the
coupling constants and of the $\varphi$ vacuum expectation value.
The connection between the mass gaps and the fundamental parameters
of the theory depend on which global symmetry is gauged, as
summarized in table \ref{tab_mass_gaps}. Moreover, all the mass gaps
can be written in terms of the fermionic gap $\Delta$. Therefore, if
one is able to build graphene with different mass gaps, for example
changing its doping and/or distortion, one can test which of the
symmetries discussed before applies to graphene, if any, simply
looking at how $\Delta_A$ and $\Delta_\Phi$ change with $\Delta$.
As an illustration, in figure \ref{fig_mass_gap_delta} we show
$\Delta_A$ and $\Delta_\Phi$ as a function of $\Delta$ when all
the coupling constants are set to unit. The figure uses arbitrary
units. Clearly, the functional behaviour distinguish between a
chiral gauge theory and a $U_b(1)$ or a $U_f(1) \otimes U_b(1)$
gauge symmetry. A $\Delta_A = 0$ for all values of $\Delta$ clearly
points towards a $U_f(1)$ gauge theory.

Of the gauge theories discussed here, the mass gap relations do not distinguish between the
two symmetries $U_b(1)$ and $U_f(1) \otimes U_b(1)$ as they provide similar types of
$\varphi - A_\mu$ interactions. However, if $U_b(1)$ does not couples directly the fermions to
the vector excitations, the $U_f(1) \otimes U_b(1)$ gauge symmetry has such a direct coupling
between electrons and gauge field.
Of course, if the strength of this coupling is extremely small, then the two theories will give exactly
the same predictions for graphene. However, if the coupling $\overline\psi \gamma^\mu \psi \, A_\mu$
is sizable enough, then the fermionic interaction with the vector excitations will give relevant contributions
to the dynamics of graphene and the two symmetries will provide different physics for  vector and
fermion excitations.

In the models discussed above, the gap in graphene is generated via spontaneous symmetry breaking.
This is not the only way of having massive particles in a theory. For example, as discussed
in \cite{chaves}, a fermionic gap can also be generated via dynamical symmetry breaking.

\section{Equations of Motion, Vortices and Flux Quantization \label{eq_movimento}}

The equations of motion associated with the various fields are derived from $\mathcal{L}$ in the usual way.
For fermions, they are given by
\begin{equation}
  i \gamma^\mu D_\mu \, \psi  ~ - ~  P (\varphi) \psi ~ - ~  P_5 (\varphi) \, \gamma_5 \, \psi~ = ~ 0 \, ,
  \label{equation_fermion}
\end{equation}
where $D_\mu = \partial_\mu + i g A_\mu$ is the covariant derivative and $A_\mu$ is the gauge field.
The corresponding equation for $\varphi$ is
\begin{eqnarray}
  & & \!\!\!\!\!\! D^\mu D_\mu \varphi ~ = ~  - \left( g_1 + g_2 \, \varphi\right)  \overline\psi \psi   +
          \left(  h_1 - i h_2 \, \varphi\right)  \overline\psi \, \gamma_5\psi
  \nonumber \\
  &  &\!\!\!\!\!\! \hspace{1.8cm}  - ~ \mu^2 \varphi   ~ -  \lambda_ 4 \left( \varphi^\dagger \varphi \right) \varphi     ~ -
  \lambda_6 \left( \varphi^\dagger \varphi \right)^2 \varphi
  \label{equation_boson}
\end{eqnarray}
with the covariant derivative given by $D_\mu = \partial_\mu + i g_\varphi A_\mu$.
The gauge field equation of motion reads
\begin{equation}
   \partial_\mu F^{\nu\mu} =
   - g \, \overline\psi \gamma^\nu \psi - i \, g_\varphi \, \varphi^\dagger \left( D^\nu \varphi \right)
                           + i \, g_\varphi  \left( D^\nu \varphi \right)^\dagger \varphi \, .
    \label{equation_photon}
\end{equation}

Equations (\ref{equation_fermion}), (\ref{equation_boson}) and (\ref{equation_photon}) are the equations
of motion derived from $\mathcal{L}$ taking into account all the possible coupling constants. In order to
study each of the gauge theories considered previously, one has to take into account the corresponding
constraints associated with the gauge group and summarized in table \ref{tab_sumario_simetrias}.

\subsection{Vortex Solutions \label{vortexsolutions}}

Let us now discuss vortex like solutions for the bosonic sector of the theory, disregarding the coupling to the
fermionic degrees of freedom. This provides a consistent solution for the field equations for the
Dirac zero modes discussed in the next section.

In this section
we will consider static solutions for equations (\ref{equation_boson}) and (\ref{equation_photon}) with
\begin{equation}
      \varphi (\vec{r}) = \varphi_0 (r) ~ e^{in\theta} \, ,
\end{equation}
and
\begin{equation}
   A^0 = 0, \qquad A^i = \epsilon^{ij} \, \partial_j a = \epsilon^{ij} \, \frac{x_j}{r} \, a^\prime(r) \, ,
   \label{A_campo}
\end{equation}
where
\begin{equation}
     a^\prime (r) = \frac{d a(r)}{dr} = b(r) \, .
\end{equation}
The equation of motion for $\varphi$ then becomes
\begin{eqnarray}
 & &   \frac{1}{r} \frac{d}{dr} \left[ r \frac{d \, \varphi_0}{dr} \right]
    ~ -  ~ \left( \frac{n}{r} \, + \,  g_\varphi \, b \right)^2  \varphi_0 =  \nonumber \\
    & &
    \qquad\qquad\qquad\qquad = \mu^2 \, \varphi_0 + \lambda_4 \, \varphi^3_0 + \lambda_6 \, \varphi^5_0
  \label{equation_varphi_vortex}
\end{eqnarray}
and the gauge field equation simplifies into
\begin{eqnarray}
    \frac{1}{r} \frac{d}{dr} \left[ r \frac{d \, b}{dr} \right]  ~ -  ~  \frac{b}{r^2}
    =  2 g_\varphi \, \left(  \frac{n}{r}   +  g_\varphi \, b \right) \varphi^2_0 \, .
  \label{equation_A_vortex}
\end{eqnarray}
Note that only for $U_A(1)$,
$U_b(1)$ and $U_f(1) \otimes U_b(1)$ gauge theories, i.e. when $g_\varphi \ne 0$,
the scalar and gauge fields are coupled. In this section we will consider only the
$U_b(1)$ and $U_f(1) \otimes U_b(1)$ symmetries.

Let us first discuss the solutions of equations
(\ref{equation_varphi_vortex}) and (\ref{equation_A_vortex}) at
small distances. For $r \ll 1$, the gauge field equation becomes
\begin{equation}
    \frac{1}{r} \frac{d}{dr} \left[ r \frac{d \, b}{dr} \right]  ~ -  ~  \frac{b}{r^2}
    =  0 \, ,
\end{equation}
provided that $\varphi_0 (r)$ is regular at the origin, whose solution is
\begin{equation}
  b(r)  = b_1 \, r + \frac{b_{-1}}{r}\, ,
  \label{b_field}
\end{equation}
where $b_1$ and $b_{-1}$ are constant of integration. If $b_{-1} = 0$, equation
(\ref{equation_varphi_vortex})
\begin{equation}
    \frac{1}{r} \frac{d}{dr} \left[ r \frac{d \, \varphi_0}{dr} \right]
   ~ -  ~  \frac{n^2}{r^2} \, \varphi_0 = 0 \ ,
    \label{varphi_short_distance}
\end{equation}
and its power law solution reads
\begin{equation}
     \varphi_0 (r) = r^{|n|} \ .
     \label{usual_vortex}
\end{equation}
On the other hand if $b_{1} = 0$, in (\ref{varphi_short_distance}) $n^2$ should be replaced by
$(n + g_\varphi b_{-1})^2$ and the corresponding scalar field solution at small $r$
is
\begin{equation}
     \varphi_0 (r) = r^{\left| n + g_\varphi b_{-1}\right|}  \, .
     \label{new_vortices}
\end{equation}
It follows that $\varphi_0 (r)$ is always regular at the origin.

At large distances, for finite energy solutions, $\varphi$ approaches a constant value, a minimum
of $V(\varphi^2_0)$, and the l.h.s of equation (\ref{equation_varphi_vortex}) vanishes.
If $\varphi_0 \ne 0$, then
\begin{equation}
   b(r) = - \frac{n}{g_\varphi \, r} \,
   \label{b_vortex}
\end{equation}
and $b(r)$ is also a solution of equation (\ref{equation_A_vortex}). Note that this solution
can be extended to full range of $r$ values, with the exception of the origin, where a delta function sets
in coming from the laplacian of $b$. On the other hand, if $\varphi_0 = 0$, i.e. for
pure graphene,  one still has a solution for $b$ as in (\ref{b_field}). However, the requirement
of finite energy demands $b_1 = 0$.

The vortex solutions include, as a particular case, the type of
configurations considered in \cite{Jackiw_Pi_2007}, where $b(r)$ is
regular and  $\varphi_0 (r) \sim r^{|n|}$, for $r \ll 1$, and, for
large $r$, $\varphi_0$ becomes constant and $b(r) \sim 1/r$. A class of vortices whose short distance behavior is given by
(\ref{new_vortices}) was found. Furthermore, the zero modes of the Dirac
equation computed in the next section require a vortex solution with
$\varphi_0 $ constant and non-vanishing and $b(r) = a^\prime (r)$
given by (\ref{b_vortex}) over all space, with the exception of the
origin as discussed previously. From the point of view of the gauge
models, the singular behavior at $r = 0$ does not raise any
conceptual problems. Indeed, we are using a continuous model to
describe graphene and the dimensions of the unit cell provide a
natural short-distance cut-off below which the model is no longer
valid or, at best, should be corrected to take into account the
crystal structure of the carbons and lattice defects.

If one takes the usual definition for the ``magnetic field",
$\vec{B} = \nabla \times \vec{A}$, it follows that $\vec B$ vanishes
at large $r$. Close to the origin, $\vec{B}$ approaches a constant
for type (\ref{usual_vortex}) solution and vanishes for
(\ref{new_vortices}) configurations. For both type of
configurations, the vortex energy
\begin{equation}
 \int d^2x \left\{ \frac{1}{2} B^2 + \left| \vec{D} \varphi \right|^2 + V(\varphi^\dagger \varphi) \right\}
\end{equation}
is finite.

Despite the vanishing of $\vec{B}$ at large distance, the ``magnetic
flux" of the vortex configuration (\ref{A_campo}) over a
sufficiently large radius closed surface is quantized. Indeed, for a
spherical surface centered at the origin,
\begin{equation}
  \Phi = \int \vec{B} \cdot \vec{dS} = \int \vec{A} \cdot \vec{dl} = - \, 2 \pi \, r \, b(r)  =  \frac{2 n \pi}{g_\varphi}
   \, ,
\end{equation}
where $n = 0, \pm 1, \dots$ is the component of the angular momentum
on an axis perpendicular to graphene plan associated with the
complex scalar field $\varphi$. Flux quantization opens the
possibility of having Bohm-Aharonov type effects in graphene without
external electromagnetic fields, where electrons are scattered by a
vector potential, which can be associated with a ``topological
defect" on the carbon structure, and acquire an extra phase. We call
the reader attention that Bohm-Aharonov phases have been observed in
suspended graphene, in association with mesoscopic deformations,
where the measured charge carriers mobility is substantial larger
than in graphene on a substrate - see \cite{Voz2010} and references
therein.

\section{Fractionalization, Dirac Equation and Zero Modes \label{fractionalizacao}}

As discussed at the beginning of the present work,
electron fractionalization is related to the normalizable zero modes of the Dirac kernel.
The presence of these zero modes opens the possibility of observation of fractional quantum Hall
effect in graphene. A nice discussion connecting the Dirac equation
zero modes with electron fractionalization can be found, for example, in \cite{Hou_et_al_2007}.

For the $U_A(1)$ gauge theory, the proof of the presence of normalizable zero modes can be found
in \cite{Jackiw_Pi_2007}. The $U_f(1)$ gauge theory includes, as a particular case, the Dirac equation
discussed by Hou, Chamon and Mudry in \cite{Hou_et_al_2007}.  Therefore, for $U_f(1)$ gauge theory
fractionalization is possible.
It remains to discuss the $U_b(1)$ and $U_f(1) \otimes U_b(1)$ gauge theories. In the following, it will
be assumed that the bosonic sector is in a static vortex configuration with $A^0 = 0$,
\begin{equation}
A^{i}=\epsilon^{ij}\partial_{j}a(r) \qquad \mbox{ and } \qquad
\varphi(\vec{r})=\varphi_{0}(r) \,  e^{in\theta}.
\end{equation}

For a general gauge field, the Dirac equation associated with the
$U_b(1)$ and $U_f(1) \otimes U_b(1)$ gauge theories is given by
\begin{eqnarray}
  & & \!\!\!\!\!\!\!\!
  \Big\{  -i \vec{\alpha} \cdot \big(\nabla - ig \vec{A}\big) \nonumber \\
   & & \, \qquad\qquad  + \,
        g_2 \left( \varphi^\dagger \varphi\right)  \beta
        \, + \,
         ih_2 \left( \varphi^\dagger \varphi\right)  \beta\gamma_5
         \Big\} \psi ~= ~ E \, \psi \, . \nonumber \\
         \label{dirac_equation}
\end{eqnarray}
 Let us define the following function
\begin{equation}
   \Delta(\vec{r}) = z \, \varphi_{0}^{2}(r) \, e^{i\alpha},\quad z=\sqrt{g_{2}^{2}+h_{2}^{2}},
\end{equation}
where
\begin{equation}
  \tan\alpha = -\frac{h_{2}}{g_{2}} \, .
\end{equation}
With the above definitions and for the Dirac spinor
\begin{equation}
  \psi = \left( \begin{array}{c}  \Psi_{+}^{b}  \\ \Psi_{+}^{a} \\ \Psi_{-}^{a} \\ \Psi_{-}^{b} \end{array} \right) \, ,
\end{equation}
the Dirac equation becomes
\begin{eqnarray}
e^{-i\theta}\left(-i\partial_{r}-\frac{\partial_{\theta}}{r}- ig a'\right)\Psi_{+}^{a}
           +\Delta(\vec{r}) \, \Psi_{-}^{a}  & = & E \, \Psi_{+}^{b} , \nonumber\\
 -e^{i\theta}\left(-i\partial_{r}+\frac{\partial_{\theta}}{r}+ ig a'\right)\Psi_{-}^{a}
           +\Delta^{*}(\vec{r}) \, \Psi_{+}^{a} & =  & E \, \Psi_{-}^{b} ,\nonumber\\
 e^{i\theta}\left(-i\partial_{r}+\frac{\partial_{\theta}}{r}+ig a'\right)\Psi_{+}^{b}
           + \Delta(\vec{r}) \, \Psi_{-}^{b} & = & E \, \Psi_{+}^{a} ,\nonumber\\
-e^{-i\theta}\left(-i\partial_{r}-\frac{\partial_{\theta}}{r}-ig a'\right)\Psi_{-}^{b}
          + \Delta^{*}(\vec{r}) \, \Psi_{+}^{b} & = & E \, \Psi_{-}^{a} , \nonumber\\
&&
\end{eqnarray}
where $a^\prime$ means the derivative with respect to $r$ of function $a(r)$.
These equations are invariant under the interchange of the two sublattices $a \leftrightarrow b$
provided that $\theta \rightarrow - \theta$ and $a^\prime \rightarrow - a^\prime$. This symmetry
generalizes the sublattice symmetry of the Dirac equation already discussed in \cite{Hou_et_al_2007}.
We proceed assuming that $\Psi^b_{\pm} = 0$. Note that, for zero modes,
given a solution of the Dirac equation in sublattice $a$, the generalized sublattice symmetry generates
another zero mode but leaving in sublattice $b$, or vice-versa. If one writes
\begin{eqnarray}
 \Psi_{+}^{a} & = & \phi_{+}(r)~ e^{i \left( m\theta + \beta_+ \right)},\nonumber\\
 \Psi_{-}^{a}  & = & \phi_{-}(r) ~ e^{i \left( k\theta + \beta_- \right)},
\end{eqnarray}
the zero mode equations become
\begin{eqnarray}
 \phi^\prime_+ + \left( \frac{m}{r} + g \, a^\prime \right) \phi_{+} + z \, \varphi_{0}^{2}(r) \, \phi_{-} & = &
                0  \label{zero_eq_1} \\
\phi^\prime_- - \left(\frac{k}{r} + g \, a^\prime \right) \phi_{-} + z \, \varphi_{0}^{2}(r) \, \phi_{+} & = & 0
\label{zero_eq_2}
\end{eqnarray}
if the following relations
\begin{equation}
  k = m - 1 \qquad \mbox{ and } \qquad \beta_+ = \frac{\pi}{2} + \alpha + \beta_-
  \label{fases}
\end{equation}
are satisfied. From equation (\ref{zero_eq_1}) one can write
\begin{equation}
  \phi_- = - \frac{1}{\Delta_0 (r)}
  \left[ \phi^\prime_+ + \left( \frac{m}{r} + g \, a^\prime \right) \phi_{+} \right] \, ,
  \label{fi_minus}
\end{equation}
where $\Delta_0 (r) = z \, \varphi^2_0 (r)$. Replacing this expression for $\phi_-$ in equation
(\ref{zero_eq_2}) one arrives at the following second order differential equation
\begin{widetext}
\begin{equation}
\phi_{+}''+\left[\frac{1}{r}-\frac{\Delta_{0}'}{\Delta_{0}}\right]\phi_{+}'  +
\left[g a''-\left(\frac{m}{r}+ga'\right)\left(\frac{k}{r}+ga'+\frac{\Delta_{0}'}{\Delta_{0}}\right)-\frac{m}{r^2}-\Delta_{0}^{2}\right]\phi_{+}=0.\nonumber
\label{zeromode_eq1}
\end{equation}
\end{widetext}
The  computation of a solution of equation (\ref{zeromode_eq1})
requires the knowledge of $\varphi_0 (r)$ and $a^\prime (r)$. Let us
look for configurations where $\varphi_0(r)$ is a non vanishing
constant, i.e. a minimum of $V(\varphi^2_0)$. Then, $\Delta^\prime_0
= 0$ and $\Delta_0 = z \, \varphi^2_0$ is also a non vanishing
constant. The equation of motion of the scalar field
(\ref{equation_varphi_vortex}) gives
\begin{equation}
  b(r) = a^\prime (r) = - \frac{n}{g_\varphi \, r} \, .
  \label{vortex_gauge}
\end{equation}
This particular gauge configuration solves the equation of motion
for the gauge field (\ref{equation_A_vortex}), except at the origin
where a Dirac delta function sets is due to the laplacian operator.
Our vortex solution requires a short-distance cut-off, which is
provided by the dimensions of the graphene unit cell or the length
scale associated with a defect. Indeed, for distances smaller than
the unit cell dimensions, the continuum description of graphene
should breakdown.

For this vortex solutions, the gauge field is linked with the
angular momenta, relative to an axis perpendicular to the graphene
sheet, of $\varphi$. Further, recall that the ``magnetic field"
associated with this type of vortex solution vanishes and,
therefore, the energy associated with the vortex also vanishes.

For a vortex with a constant $\varphi_0$, equation
(\ref{zeromode_eq1}) simplifies to
\begin{equation}
\phi_{+}'' + \frac{1}{r} \, \phi_{+}'  +
 \left[ -\frac{1}{r^2} \left( m - \frac{g}{g_\varphi} n \right)^2 -\Delta_{0}^{2}\right]\phi_{+}=0.\nonumber
\label{zeromode_eq2}
\end{equation}
The solutions of this equation are the modified Bessel functions $I$ and $K$ of argument $\Delta_0 r$ for
particular combinations of the angular momenta associated with $\varphi$, $\phi_+$ and $\phi_-$
-- see appendix \ref{Solucoes_Dirac} for details.

The gauge model has, at least, one normalizable zero mode state of
the Dirac equation. Therefore, the $U_b(1)$ and $U_f(1) \otimes
U_b(1)$ gauge models can accommodate electron fractionalization
without the presence of external electromagnetic fields. More, given
the vortex solution and relation (\ref{vortex_gauge}), besides
fractionalization, the gauge models also incorporates flux
quantization associated with the ``magnetic field". The phase-shifts
coming from the Bohm-Aharonov effect associated with the vortex are
connected with the component of the angular momenta of $\varphi$
along an axis perpendicular to the graphene layer.

\subsection{Zero modes and Conformal Invariance Breakdown}

The Sturm-Liouville form of the zero mode equation is found by
setting $\phi_{+} = F / \sqrt{z}$, with $z=\Delta_0\; r$, in
equation (\ref{zeromode_eq2}), which becomes
\begin{equation}
F'' + \frac{1}{z^2}\left[\frac{1}{4} - \nu^2 \right] F= F \ ,
\label{stleq}
\end{equation}
where $\nu^2 = \left( m - \frac{g}{g_\varphi} n \right)^2 $. This
equation is equivalent to a one-dimensional Schr\"odinger eigenvalue
problem with a potential $1/z^2$ which expresses invariance under
scale change.  The conformal symmetry is broken by the ultraviolet
physics associated with defects/unity cell scales. Given that
$\nu^2\geq 0$, the potential strength for the Schr\"odinger problem
is above the Breitenlohner-Freedman bound \cite{BFB} and the
corresponding quantum mechanical model is free from
instabilities, i.e. the zero mode state does not collapse. In
other contexts, for example in ultracold atomic physics, the violation
of this bound gives rise to the Efimov effect\cite{braaten}.

We observe the connection between the fermionic Sturm-Liouville
equation and the dynamics of fermion fields in a supergravity
Anti-de-Sitter (AdS) background -- see, for example, \cite{review}.
In this description, the metric embodies conformal invariance leading to
a $1/r^2$ potential associated, in our case, with the vortex solutions.
The mass term of the fermionic field in the corresponding
supergravity action contains the factor $\nu^2$ (see e.g.
\cite{forkel}). Within this framework, the required short-range
regularization (see the appendix) could be performed at the expense
of introducing a dilaton field coupled to gravity, deforming the AdS
metric \cite{wayne,wayne2010}. This suggests that the Maldacena conjecture of
the AdS/CFT (conformal field theory) duality
\cite{maldacena,witten}, may well provide fresh insights to graphene
physics. For example, the vector mass gap could
be a consequence of breaking exact symmetries in holographic
10-dimensional backgrounds, encoding mass gaps for the fermionic
field and vector fluctuations \cite{Mass}.

\section{Results and Conclusions \label{fim}}

In this paper we discuss gauge theories for graphene. The building of the gauge models starts assuming that
graphene dynamics can be described by fermion fields together with a complex scalar field $\varphi$. The field
$\varphi$ resumes the self-interaction of the carbon background and the mean fermionic self interaction.
After exploring the global symmetry of the most general lagrangian, excluding derivative-like couplings,
the corresponding gauge models are investigated.

The gauge models are compatible with a gap for fermion, vector and
scalar fields. The mass gaps are generated via an Higgs mechanism.
Further, the mass gaps associated with each gauge model are
connected in different ways, which opens for the possibility of
experimentally distinguish between the models. Indeed, as claimed in
section \ref{seccao_mass_gap}, for example, changing the
concentration of impurities in graphene, one can change the
fermionic gap and check how the scalar and vector gaps adjust and,
in this way, check which of the gauge models, if any, reproduce the
graphene results. Furthermore, in what concerns the fermionic gap,
within the models considered here, the mass gap is a dynamical
quantity associated with the field $\varphi$.

The gauge models have finite energy vortex solutions. Therefore,
phenomena like the flux quantization of the ``magnetic field", in
association with topological defects of the carbon structure, and/or
Bohm--Aharonov type effects become possible within the description
of graphene by gauge models. Several types of vortex solutions where
discussed and, in general, the gauge field is linked with component
of the angular momenta, along an axis perpendicular to the graphene
plan, of $\varphi$. For this type of solution the phases associated
with Bohm--Aharonov type effects are a measure of the $\varphi$
angular momenta.

Finally, we have investigated the two dimensional Dirac equation for
the vortex solutions. A generalization of the sublattice symmetry
was discussed and we showed that all gauge models have normalizable
zero modes together with a non-vanishing fermionc gap.
Fractionalization is then possible in all the gauge models
considered here. Within this theoretical background one expects that
fractional quantum Hall effect can take place in graphene in
connection with the zero mode solutions, even when there are no
external electromagnetic fields.
See also the discussion at the end of section \ref{introducao}.

The models investigated are potentially useful to describe graphene.
Indeed, they unify, under the same dynamical principle, several
features of the graphene and predict others. The models have
multiple parameters whose values should be found by reproducing
experimentally known  graphene properties. We are currently engaged
in performing such an investigation and will report the results
elsewhere.

\appendix

\section{Zero Modes of the Dirac Equation \label{Solucoes_Dirac}}

Let us discuss the solutions of equation (\ref{zeromode_eq2}),
\begin{equation}
\phi_{+}'' + \frac{1}{r} \, \phi_{+}'  +
 \left[ -\frac{1}{r^2} \left( m - \frac{g}{g_\varphi} n \right)^2 -\Delta_{0}^{2}\right]\phi_{+}=0 \, .
\end{equation}
Introducing the adimensional distance $z = \Delta_0 \, r$, after multiplying this equation by $z^2$ one gets the
following differential equation
\begin{equation}
z^2 \, \phi_{+}'' + z \, \phi_{+}'  -
 \left[ \left( m - \frac{g}{g_\varphi} n \right)^2 + 1\right]\phi_{+}=0
\end{equation}
whose solutions are the modified Bessel functions $I_{\pm \nu} (z)$ and $K_\nu (z)$, see \cite{Abramowitz} for definitions,
where
\begin{equation}
  \nu^2 = \left( m - \frac{g}{g_\varphi} n \right)^2 \, .
\end{equation}
Note that $\nu$ can be a real number. Of $I_{\pm \nu} (z)$ and $K_\nu (z)$ functions, only the last one tends to zero as
$r \rightarrow + \infty$. Indeed, in this limit
\begin{equation}
  K_\nu (z) ~ = ~ \sqrt{\frac{\pi}{2 z} } ~ e^{-z} \, ,
\end{equation}
with its first derivative having a similar functional behavior. Therefore, modulo the behavior for small $r$, in principle,
setting $\phi_+$ proportional to $K_\nu$ the differential equation (\ref{zeromode_eq2}) is solved and the spinor
is normalizable.

For small $r$
\begin{equation}
  K_\nu (z) ~ = ~ \frac{1}{2} \Gamma ( \nu ) \left( \frac{1}{2} z \right)^{- \nu}
  \label{K_nu_small_r}
\end{equation}
and the spinor diverge.

The divergence of $\phi_+$ can be resolved as described in
\cite{Landau} in connection with the potential $ - \beta / r^2$ with
$\beta > 0$. A short distance cut-off $r_0$ is introduced and the
``potential" is replaced by its value at $r_0$. However, if $\phi_+$
can be made regular at the origin, $\phi_-$ being given by
(\ref{fi_minus}) will diverge near the origin. Due to this short
distance divergence, $\phi_-$ is not normalizable unless the model
has a minimal distance beyond which the continuum description of
graphene no longer makes sense. The dimensions of the unit cell
provide such an infrared cut-off. Remember that, in graphene, the
carbon atoms are separated by $a \approx 1.42 \, \AA$ and for $r <
a$ one can set $\phi_\pm (r) \approx \phi_\pm (a)$ and have a
continuous and normalizable spinor.

We call the reader attention that the introduction of a short distance, i.e. ultraviolet,
cut-off for the fermions fields does not change the results of  section
\ref{vortexsolutions}.
Indeed, the Dirac zero modes give no contribution to the equations of motion
associated with the vortex solution of section  \ref{vortexsolutions}.
In this sense, the flux quantization
of the gauge field and the Bohm-Aharonov effect discussed there,
i.e. the topological properties of the vortex solution,  are independent
of the Dirac spinors.
On the other hand, the requirement that the Dirac spinors
are normalizable is a necessary condition to have electron fractionalization;
see, for example, the discussion in the
work of Hou \textit{et al} \cite{Hou_et_al_2007}.

The singular behavior at $r = 0$ seem to be an indication of a
"hole" at the center of the unit cell. This "hole" is a topological obstruction
and is at the origin of the topological properties of the model analyzed in the
present work.

\section*{Acknowledgements}

The authors acknowledge financial support from the Brazilian
agencies FAPESP (Funda\c c\~ao de Amparo \`a Pesquisa do Estado de
S\~ao Paulo) and CNPq (Conselho Nacional de Desenvolvimento
Cient\'ifico e Tecnol\'ogico).


\end{document}